\documentclass[sigconf,natbib=false]{acmart}
\AtBeginDocument{%
  \providecommand\BibTeX{{%
    \normalfont B\kern-0.5em{\scshape i\kern-0.25em b}\kern-0.8em\TeX}}}

\RequirePackage[
  style=ACM-Reference-Format,
  sortcites,
  doi=false,
  url=false,
  isbn=false,
  maxbibnames=6,
  natbib=true,
  backend=bibtex
  ]{biblatex}

\usepackage{threeparttable}
\usepackage{multirow} 
\usepackage{makecell}
\usepackage{url}
\usepackage{inputenc}

\renewcommand\footnotetextcopyrightpermission[1]{}

\begin{document}
\pagestyle{plain}

\title{
{\vspace{-1cm}\small This is the author's version. To appear at the 41st IEEE/ACM International Conference on Computer-Aided Design (ICCAD '22), October 30-November 3, 2022, San Diego, CA, USA. https://doi.org/10.1145/3508352.3561105}\\\vspace{-0.5\baselineskip}\rule{\textwidth}{0.4pt}\vspace{1cm}
Approximate Computing and the\\
Efficient Machine Learning Expedition}
\subtitle{(Invited Paper)\vspace{3ex}}

\author{J\"org Henkel}
\email{henkel@kit.edu}
\orcid{https://orcid.org/0000-0001-9602-2922}
\affiliation{%
  \institution{Karlsruhe Institute of Technology}
  \city{Karlsruhe}
  \country{Germany}}

\author{Hai Li}
\email{hai.li@duke.edu}
\orcid{https://orcid.org/0000-0003-3228-6544}
\affiliation{%
  \institution{Duke University}
  \city{Durham}
  \country{USA}}
  
\author{Anand Raghunathan}
\email{raghunathan@purdue.edu}
\orcid{https://orcid.org/0000-0002-4624-564X}
\affiliation{%
  \institution{Purdue University}
  \city{West Lafayette}
  \country{USA}}

\author{Mehdi B. Tahoori}
\email{mehdi.tahoori@kit.edu}
\orcid{https://orcid.org/0000-0002-8829-5610}
\affiliation{%
  \institution{Karlsruhe Institute of Technology}
  \city{Karlsruhe}
  \country{Germany}}

\author{Swagath Venkataramani}
\email{swagath.venkataramani@ibm.com}
\orcid{https://orcid.org/0000-0002-0470-6364}
\affiliation{%
  \institution{IBM Research}
  \city{Yorktown Heights}
  \country{USA}}
  
\author{Xiaoxuan Yang}
\email{xy92@duke.edu}
\orcid{https://orcid.org/0000-0002-2553-2631}
\affiliation{%
  \institution{Duke University}
  \city{Durham}
  \country{USA}}

\author{Georgios Zervakis}
\email{georgios.zervakis@kit.edu}
\orcid{https://orcid.org/0000-0001-8110-7122}
\affiliation{%
  \institution{Karlsruhe Institute of Technology}
  \city{Karlsruhe}
  \country{Germany}}
\additionalaffiliation{%
  \institution{University of Patras}
  \city{Patras}
  \country{Greece}
}

\begin{abstract}
Approximate computing (AxC) has been long accepted as a design alternative for efficient system implementation at the cost of relaxed accuracy requirements.
Despite the AxC research activities in various application domains, AxC thrived the past decade when it was applied in Machine Learning (ML).
The by definition approximate notion of ML models but also the increased computational overheads associated with ML applications–that were effectively mitigated by corresponding approximations–led to a perfect matching and a fruitful synergy.
AxC for AI/ML has transcended beyond academic prototypes.
In this work, we enlighten the synergistic nature of AxC and ML and elucidate the impact of AxC in designing efficient ML systems.
To that end, we present an overview and taxonomy of AxC for ML and use two descriptive application scenarios to demonstrate how AxC boosts the efficiency of ML systems. 

\end{abstract}



\maketitle

\section{Introduction}

The failure of Dennard scaling led to the so-called dark silicon problem~\cite{Henkel:DAC2015:darksil} and computer designers were forced to explore radical new approaches to sustain and further improve the efficiency of our computing systems.
Several groundbreaking novelties came from the field of computing.
Among these newly established computing paradigms, over the past decade, a phenomenal boom is observed in Approximate Computing (AxC)~\cite{Venkataramani:DAC2015:axc} research.
Approximate computing refers to techniques that exploit the inherent error resilience of several applications to achieve improvements in efficiency (e.g., energy and performance) at all layers of the computing stack~\cite{Venkataramani:DAC2015:axc}.
For example, prior analysis on a benchmark suite of 12 recognition, mining and search applications showed that 83\% of the runtime is spent in tasks that are amenable to approximation~\cite{Chippa:DAC2013,Venkataramani:DAC2015:axc}.
The origins of approximate computing (AxC) can be traced back to various fields including computer arithmetic (floating point representation)~\cite{AxNN}, arithmetic units (adders~\cite{Shafique:DAC2015:gear} and multipliers~\cite{multperforation}), digital signal processing (filter design)~\cite{Gupta:TCAD2013}, algorithms (approximation algorithms)~\cite{ScalableEffort}, and networking (best-effort packet delivery)~\cite{Chakradhar:DAC2010}. 

However, the field really took off in the past decade when AxC intersected with machine learning (ML)~\cite{ScalableEffort,axdnnsurv,swagath2020}, which arguably provides the ideal workload for AxC due to several factors. Many ML tasks at some level reduce to a problem of function approximation, where the function is incompletely specified, allowing ample scope for AxC. Further, the very process used to create models (training) can be re-purposed to recover from any adverse effects of approximations. In particular, AxC and deep neural networks (DNNs)--in which the latest improvements in the accuracy came at the cost of a great increase in computational requirements-- formed a perfect match~\cite{axdnn:DATE2018,convar,multiplierless,axweight}. Due to this synergy, the field of AxC witnessed significant interest and growth, and expanded into new levels of the computing stack such as approximate hardware.
AxC for ML (e.g., quantization~\cite{2bitquant}, pruning~\cite{structuredsparsity}, relaxed synchronization~\cite{relaxedsync}) has already been adopted in practice and is a key enabler of the AI hardware roadmap for many companies (e.g., ultra-low precision~\cite{rapid} and analog hardware~\cite{memristorxbar}).

In this work, we first present an end-to-end view of designing approximate computing systems for ML, spanning algorithms to circuits, and discuss the efficiency benefits accrued from AxC, highlighting also current commercial hardware and software that employ AxC.
Next, we present two notable examples of approximate ML systems.
Section~\ref{sec:pim} focuses on in-memory-computing-based (IMC-based) approximation for the state-of-the-art Transformer models.
Transformer has become a prominent neural network model for Neural Language Processing (NLP) applications, with outstanding results in neural machine translation, entity recognition, etc.
Section~\ref{sec:printed} discusses the exploitation of approximate computing as a key enabler for the realization of ultra-resource constrained ML circuits and specifically of battery powered printed ML classifiers.
Printed electronics form an extreme use-case of embedded ML and pose a very promising solution to enable computing in application domains untouchable by silicon-based systems.
Concluding, we discuss some key challenges that need to be addressed to enable further growth and adoption of AxC, especially in ML.


\section{AxC for ML: A Taxonomy}\label{sec:overview}
  
\begin{figure}[t!]
\centering
\includegraphics[width=\linewidth]{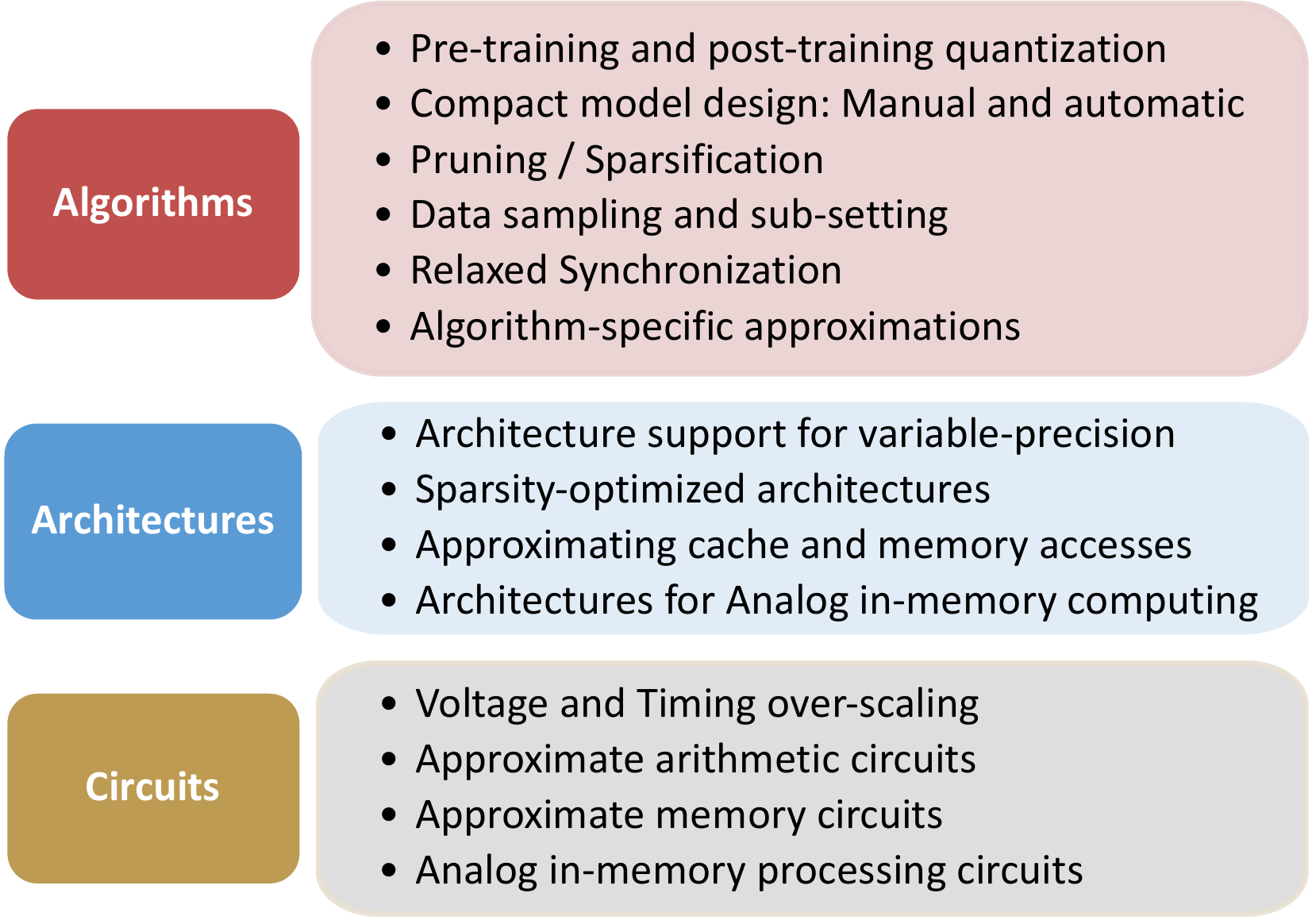}
\caption{Taxonomy of Approximate Computing techniques for AI/ML across different levels of the compute stack}
\label{fig:axc-taxonomy}
\vspace{-4ex}
\end{figure}

Recognizing the opportunity for significant improvements in  performance and compute efficiency, developing Approximate Computing (AxC) techniques for AI/ML applications has been an active area of research in the past decade. The techniques spanned the entire computing stack from algorithms to circuits. Figure~\ref{fig:axc-taxonomy} shows a taxonomy of the various approximate computing techniques targeting AI/ML applications. Not surprisingly, the most successful techniques are inherently cross-layered \emph{i.e.,} multiple layers of the compute stack are co-designed to achieve maximum benefits from the approximations, while minimizing their impact on end-application output quality~\cite{ScalableEffort}. The following subsections describe the AxC design techniques at each level of the compute stack in more detail.

\subsection{Algorithmic Approximations for AI/ML}
The goal of approximation techniques at the algorithm level is to identify the degree to which computations and data of the application can be approximated thereby driving down the compute and memory requirements and heal the impact of approximations on end-application quality. The majority of algorithmic approximations require careful co-design with the hardware layers of the compute stack to maximally exploit the benefits, while some approximations can be realized without specialized hardware support.

Some of the popular approximation techniques for AI/ML applications are listed below:

{\bf \noindent Quantization or Precision Scaling.} Amongst the different approximation techniques, quantization has proven to be most successful technique widely adopted by the industry in the context of both training and inference of AI/ML models. At a high level, quantization involves scaling or reducing the number of bits used to represent the weights and activations. Reducing the numeric bit-precision super-linearly improves compute efficiency (for example, a 16-bit multiplier is roughly $\sim$4 times lower cost than a 32-bit multiplier), and linearly reduces memory footprint and bandwidth. Further, it preserves the regularity in the compute patterns prevalent in AI/ML models, which makes it amenable to be integrated within both general-purpose and accelerator-based designs.

The key challenge in quantization is to ensure that the loss in numeric precision does not impact functional accuracy. In the context of training, research efforts have successfully explored different 16-bit floating point (FP16) variants \emph{viz.} IEEE-FP16 (1-5-10), BFloat (1-8-7)~\cite{Google:2019} and DLFloat (1-6-9)~\cite{Agarwal:2019}. More recently, a new Hybrid-FP8 (HFP8) data format~\cite{sun2019hybrid}, which involves combining two
different FP8 formats---(1,4,3) for activations and weights, and (1,5,2)
for errors---was introduced to lower training requirements even further.

In the context of inference, the use of fixed-point (FxP) number representation is quite prevalent. There are two prevailing approaches to minimize the impact of using reduced-precision FxP for inference---post-training quantization (PTQ)
and quantization-aware training (QAT). Post-training quantization involves directly turning a model trained with floating-point representation into a fixed-point model~\cite{zhao2019improving}. The loss in accuracy due to direct quantization can be minimized by re-calibrating batchnorm statistics. However, state-of-the-art PTQ techniques lose considerable model accuracy especially below 8-bits. Quantization-aware training (QAT) methods, first introduced in AxNN~\cite{AxNN}, retrain the model by reflecting the effect of quantization within the training process so that the weights can be suitably adapted to heal the impact of quantization. Popular QAT methods such as Parameterized Activation Clipping (PACT) and statistics-aware weight binning (SAWB) achieve state-of-the-art classification accuracy with INT4 precision, and incur only a small accuracy loss with INT2 across broad range of domains~\cite{Choi:2018}. 

It is crucial to note that quantization is typically applied only to convolution (CONV) and matrix-multiply (GEMM) operations in an AI/ML model. While CONV/GEMM constitute the bulk of computations, other auxiliary operations for activation, pooling and normalization are typically performed in FP16 format. Also, in some cases, each layer of the AI/ML model is quantized to a different bit-precision.

\vspace{3pt}
{\bf \noindent Compact Model Design: Manual And Automatic.} There is an urgent need to deploy AI/ML models in extremely resource-constrained mobile/IoT devices. One approach to facilitate this is to design compact AI/ML models that achieve near state-of-the-art accuracies as their large-scale counterparts. Popular neural network topologies such as inverted-bottleneck structures, depth-wise seperable convolutions, group convolutions, significantly reduce the model size and consequently the number of computations. Recently, automatic methods to train compact models such as knowledge distillation~\cite{hinton2015distilling} and Neural Architecture Search (NAS)~\cite{Elsken:2019} are gaining popularity.

\vspace{3pt}
{\bf \noindent Pruning / Sparsification and Compression.} Another increasingly popular approach to reduce model size is \emph{pruning}. It forces some weight values in a neural network to zero, thereby eliminating connections between associated neurons. Pruning sparsifies the weight tensors, which are then compactly stored and accessed using sparse representations (\emph{e.g.,} index-value pairs). One of the key challenges in pruning is that it makes the compute and memory access pattern irregular and hence does not lend itself well into hardware acceleration. Recent approaches perform pruning in a structured fashion that preserves regularity~\cite{structuredsparsity}. One popular example is \emph{fine-grained block sparsity}, wherein the weight tensor is split into equi-sized element blocks and each block is restricted to have a fixed percentage of non-zero elements~\cite{Nvidia:2021}. Akin to quantization, pruning is performed at training time in order to heal its impact on accuracy. 

In the context of training, which is typically performed in a distributed system, one of the key bottlenecks to overall throughput is off-chip communication. Training is parallelized by splitting the minibatch inputs across multiple learners and their respective weight gradients need to be accumulated to compute the updated weight value. To reduce off-chip communication cost, the weight gradient values are compressed using a number of interesting algorithms~\cite{Chen:2017} that have achieved over $>$100$\times$ compression rates without loosing model accuracy.

\vspace{3pt}
{\bf \noindent Data Sampling and sub-setting.} Data sampling is a technique where only a subset of elements of a tensor is used for computations. Sampling reduces the overall tensor footprint, which enables it to be contained within the lower levels of the memory hierarchy, alleviating performance bottlenecks due to memory latency and bandwidth limitations. Sampling techniques~\cite{Kim:2019} typically exploit \emph{value similarity} between data-elements present in a region. For example, in real-world images, adjacent pixels take similar values. In some cases, sampling can result in some of the computations to be redundant, which can be eliminated.

\vspace{3pt}
{\bf \noindent Relaxed Synchronization.} Relaxed synchronization is a technique which is applied in the context of distributed training. Broadly, training algorithms require the learners to be synchronized at the end of each minibatch iteration so that model weights are updated using the weight gradients from each learner. Relaxed synchronization allows the learners to proceed semi-synchronously or even asynchronously \emph{i.e.,} fast learners do not wait for slow learners to complete but proceed immediately to the next iteration~\cite{Zhang:2016}. The key trade-off is to ensure that this does not result in requiring additional training epochs for model convergence. 

\subsection{Approximate AI/ML Architectures}
AI/ML applications find utility under a variety of deployment scenarios from cloud to the edge, with differing latency, throughput, and energy requirements. Hardware specialization/acceleration is regarded critical to meet the computational demands of AI/ML models, while a bulk of AI/ML applications are still executed on general-purpose CPUs owing to cost and form-factor constraints. Hence, approximations identified at the algorithm level have to judiciously incorporated within a spectrum of compute architectures from general-purpose processors to application-specific designs to maximally exploit the benefits from approximate computing. 

One of the key tenets for approximate architecture design is that approximable components should dominate the overall energy consumption. Inherent control front-ends for instruction fetch/decode \emph{etc.} needs to be executed in an accurate manner. Hardware accelerators for AI/ML contain arrays of processing elements which exploit the abundant data parallelism available in these applications with minimal control components. The execution engines dominate their overall energy consumption, and hence AxC computing techniques can be effectively employed to scale their performance and energy. Some of the design techniques at the architecture level include:

\vspace{3pt}
{\bf \noindent Architecture Support for Variable Precision.} As mentioned earlier, the precision to which the data elements can be quantized varies widely across application domains, deployment scenarios and across layers of the model. This necessitates AI/ML architectures to design execution engines with mixed-precision support, whose accuracy levels are directly controlled by the software~\cite{Venkataramani:2013}. A key challenge in embodying mixed-precision support lies in identifying the right degree of logic sharing across the different precisions. Too much sharing would lead to poor energy efficiency at any given precision, whereas a completely decoupled execution pipeline for each precision would lead to significant area and leakage power overheads. A variable-precision AI accelerator design is the RaPiD architecture~\cite{rapid}, fabricated at 7nm technology, that supports precisions ranging from 16-bit floating point to 2-bit fixed point, scaling peak compute capability from 8 TFLOPs to $\sim$200 TOPS.

\vspace{3pt}
{\bf \noindent Sparsity-optimized Architectures.} The model pruning approximation triggers the need for architectures that can exploit sparsity in data to benefit performance and energy. Sparsity also naturally occurs in activations due to the use of ReLU activation function, particularly in the vision domain. Sparse accelerator architectures~\cite{Han:2016, Pal:2018} under a variety of scenarios---sparsity in one or both operands (weights and/or activations), support for structured \emph{vs.} unstructured sparsity---have been explored in literature. The interplay between sparsity and quantization is an open challenge. The overheads due to sparse execution are amplified as the execution engines shrink with quantization favoring dense computation arrays. In addition to specialized architectures, sparsity support has also been explored in the context of general-purpose processors~\cite{Sen:2019}. The key idea is to dynamically pre-detect and skip a set of future instructions that are rendered ineffectual due to sparsity. 

\vspace{3pt}
{\bf \noindent Approximating Cache and Memory Accesses.} The memory sub-system is often a critical bottleneck to performance and energy efficiency. Techniques such as quantization and pruning naturally benefit both compute and memory, but more in favor of compute due to super-linear reduction in multiple cost and overheads of sparse index representations. This leaves the memory sub-system still as the bottleneck. Addressing this challenge, specific approximation techniques targeting the memory sub-system have been proposed. These include reducing DRAM refresh rates~\cite{Liu:2011}, speculating on the results of loads~\cite{Miguel:2014}, redirecting accesses to data already available at a memory level~\cite{Kim:2019}, among others. 

\vspace{3pt}
{\bf \noindent Architectures for Analog In-Memory Computing.} In-memory computing is an exciting class of architectures actively explored in recent years for AI/ML applications. It attempts to overcome the classical von-neumann bottleneck inherent in any dataflow architecture that transfer data between the compute and memory elements. This is achieved by deeply embedding the compute with the memory array itself. In-memory computing is generally based on analog processing and has been explored in the context of SRAM~\cite{Kang:2018} and non-voltatile memories such as ReRAM, MRAM, or PCM~\cite{chakraborty2020resistive}. Section~\ref{sec:pim} presents a detailed case study outlining an in-memory accelerator architecture for Transformer models.

\subsection{Approximation Circuit Design}
Approximate circuits are the basic building block of any approximate computing system. Techniques such as quantization and other number representations for AI/ML rely on efficient approximate circuit design to leverage maximum benefits. From the bottom-up, approximations at the circuit-level can be classified as: (i) logic approximation, wherein the logic functionality of a circuit is modified slightly to realize a disproportionately efficient implementation~\cite{Venkataramani:2012}, (ii) timing approximation, wherein the circuit is designed and operated at an over-scaled voltage-frequency point that benefits efficiency while introducing modest timing errors~\cite{ScalableEffort}. Prior research also focused on designing approximate memory circuits whose access accuracy can be scaled dynamically at execution time~\cite{Ranjan:2015}. 

In summary, the special synergy between AxC and AI/ML has enabled unprecedented levels of compute efficiency through systematic co-design across the compute stack. AxC for AI/ML has transcended beyond academic prototypes. Commercial products---CPUs, GPUs and accelerators---targeting AI/ML applications embody AxC techniques such as low precision and fine-grained structured sparsity. AxC forms an integral part of the AI/ML roadmap for many companies. 


\section{Transformers on In-memory Accelerators Use Case}
\label{sec:pim}

In-memory computing~(IMC) designs can eliminate the cost of data transfer since the computation is performed within the memory device. And in-memory computing can be efficiently implemented using emerging technologies. For instance, resistive random-access memory~(ReRAM) is one of the most promising emerging technologies and has potentials to perform vector-matrix multiplications compared with CMOS accelerators~\cite{yang2022research}. 
Prior ReRAM-based processing-in-memory~(PIM) works have demonstrated their potentials in improving the efficiency of neural networks training and inference, such as Brain-State-in-a-Box~(BSB) model, deep convolutional neural network, and generative adversarial networks (GAN)~\cite{DBLP:conf/hpca/SongQ0C17,DBLP:conf/aspdac/ChenSC18,DBLP:journals/tnn/HuLCWRL14}. Note that these in-memory computing designs are essential approximate computing cases. In this section, we discuss the general approximate in-memory computing and approximate in-memory computing in Transformer-based models.

\subsection{General Approximate IMC}
The analog basic of in-memory computing matches the concept of approximate computing. Take the ReRAM crossbar-based computation as an example, the weights are programmed as the conductance of ReRAM cells. Meanwhile, the voltage, which encodes the input information, is supplied to the word lines. Here, digital-to-analog and analog-to-digital converters are utilized to assist the signal conversion between analog and digital domains. When converting the weight matrix and input into the conductance matrix and voltage, they use fixed-point encoding with a limited bit precision to represent computation elements~\cite{yang2021multi}. According to Kirchhoff's circuit law and Ohm's law, the output currents accumulated through the bit lines represent the matrix-vector multiplication results. In this case, the analog values stored on the device or supplied to it approximate the values to be computed, and the analog results generated by the crossbar approximate the final results. Adjusting the data representation precision can tradeoff the accuracy result and computing efficiency~\cite{DBLP:conf/hpca/SongQ0C17}. However, due to the inherent error tolerant ability of neural networks, limited-precision data representation will not degrade the accuracy performance~\cite{hu2016dot}.

Furthermore, the nonidealities in the IMC system lead the computation to be inaccurate~\cite{chakraborty2020resistive}. In the programming stage, the voltage applied to the electrode layers changes the resistance status of the ReRAM cell. To be more specific, when a voltage is supplied to the ReRAM cell, a conductive filament composed of oxygen vacancies may form. Note that the randomness of generating oxygen vacancies is the primary cause of process variation. The geometry of oxygen vacancies is unexpected under the same level of voltage and equivalent circumstances, resulting in a variance in resistance levels between cycles. 

There are statistical models to mimic the influence of IMC nonidealities. Variation in the resistance of the crossbar array obeys the lognormal distribution~\cite{lee2012multi,hsu2015study}, and the reason for this lognormal distribution is the Gaussian distribution of the average gap distances~\cite{yu2012switching}. Aside from the variation, the IMC with ReRAM also suffers from thermal noise, shot noise, and random telegraph noise~(RTN)~\cite{feinberg2018making}. For detailed noise distributions and analysis, we refer interested readers to ~\cite{yang2021multi,he2019noise}. 

\begin{figure}[t]
\centering
\includegraphics[width=\linewidth]{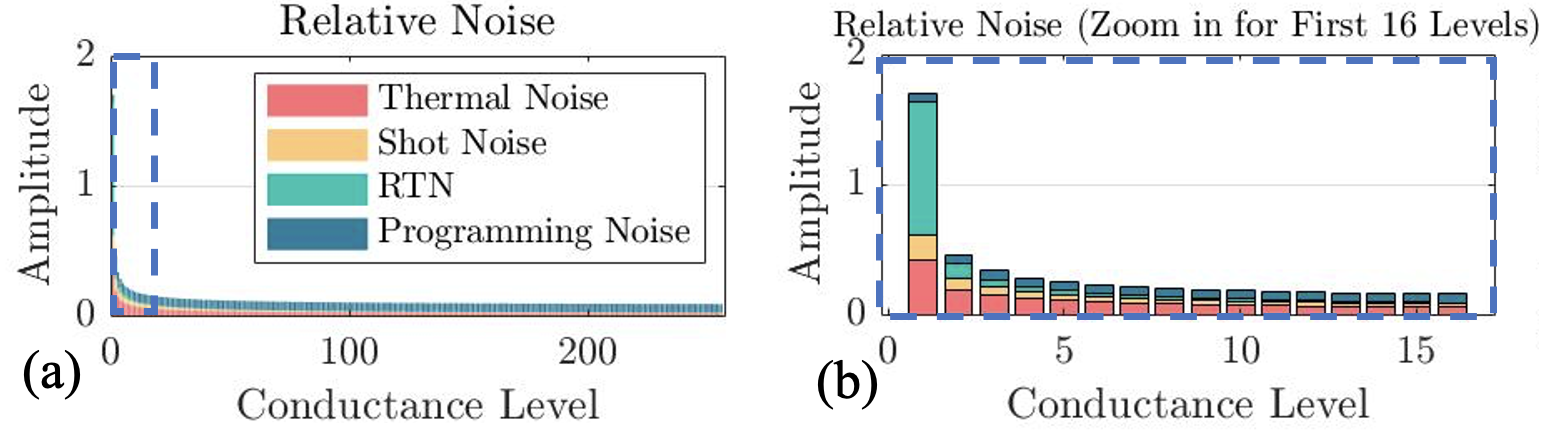}
\caption{The distributions of stochastic noise for 8-bit ReRAM cells under $Freq=500\,\mathrm{MHz}$ and $T=350\,\mathrm{K}$: (a) in a relative scale,
(b) zoomed-in version for relative noise.}
\label{fig:noise}
\vspace{-4ex}
\end{figure}

The stochastic noise for 8-bit ReRAM cell is shown in Figure~\ref{fig:noise}. The relative noise is calculated with the noise amplitude divided by the absolute conductance in each cell. In Figure~\ref{fig:noise}(a), the relative noise of small conductance level is greater than that of large conductance level, which indicates that the stochastic noise has a larger impact on small conductance values. As shown in Figure~\ref{fig:noise}(b), RTN and thermal noise dominate the stochastic noise at small conductance levels. As the thermal noise increases with the temperature~($T$) and operating frequency~($Freq$), the influence of stochastic noise will be amplified when the IMC system is running at high temperature and high frequency.

The cell variation can accumulate through the accumulation via each column, and thus the variation of different computing results may overlap~\cite{chen201865nm}. This overlapping variation problem will be severe if either the number of word lines or the number of low-resistance cells gets larger~\cite{kang2020minimizing}. Therefore, the possible approaches to minimize the calculation variation are reducing the number of word lines calculated in each cycle and increasing the number of high-resistance cells within a specific rounding distance. And the IMC-based approximate computing by adjusting the number of word lines will tradeoff between performance and accuracy.

To tackle the problem of variation and noise in IMC, circuit-level designs with feedback loops can effectively reduce the variation influence. For instance, a closed-loop circuit design utilizes the inferencing results to stabilize the conductance values on the device~\cite{yan2017closed}. Under this circumstance, the conductance values programmed onto the device are slightly different from the pure-software training weights because these values are determined with the integration of the real-device variation. Following the hardware-aware strategy, the noise injection adaption framework adjusts the conductance value by taking randomly sampled noise into account~\cite{he2019noise}. Similarly, the stochastic-noise-aware training method can mitigate the intrinsic noise in IMC system and improve the inferencing accuracy under high-frequency and high-temperature settings~\cite{yang2021multi}.

\subsection{Approximate IMC in Transformer-based Models}

\begin{figure}[t]
\centering
\includegraphics[width=\linewidth]{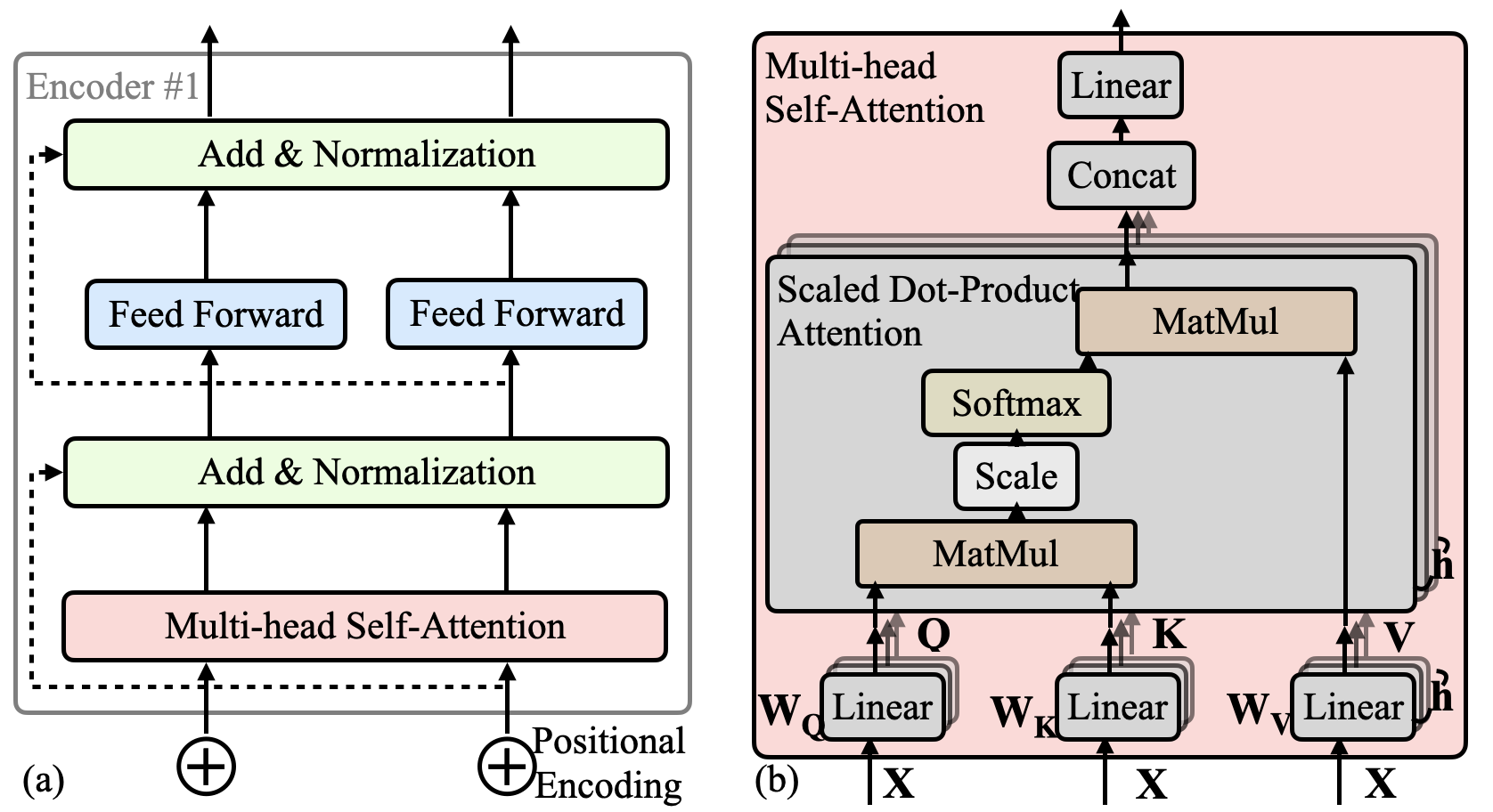}
\caption{Transformer model structure: (a) encoder structure, (b) 
multi-head self-attention module.}
\label{fig:tran}
\vspace{-4ex}
\end{figure}

Transformer has become a prominent deep neural network model with outstanding results in neural machine translation, entity recognition, etc~\cite{vaswani2017attention}. Transformer-based pre-trained models, such as Bert~\cite{devlin2018bert}, have served as the backbone of NLP applications. 
The Transformer model consists of an encoder stack and a decoder stack, which shares similar self-attention-based structure. Figure~\ref{fig:tran} shows essential components in Transformer model. The major challenge of utilizing IMC for efficient Transformer model is that the multi-head self-attention block brings computation and programming dependency and severe latency overhead. In prior works of improving the computing efficiency with PIM inference, they utilize the weight-stationary approach and don't have to reprogram the weights during the process~\cite{shafiee2016isaac}. However, the self-attention block requires the computation of two intermediate results, such as the Query $\mathcal{Q}$ and Key $\mathcal{K}$, as shown in Figure~\ref{fig:tran}(b). Therefore, performing the matrix-matrix multiplication~(MatMul) of these intermediate results with PIM will involve the programming of one intermediate result matrix onto the memory device. The computation phase has to be paused until the row-by-row matrix programming is completed. Hence, the computation efficiency will be compromised due to the compute-after-write dependency~\cite{yang2020retransformer}. 

Therefore, it is necessary to design IMC architecture to improve the computation efficiency of multi-head self-attention block.
ReTransformer is a ReRAM-based IMC design for Transformer acceleration~\cite{yang2020retransformer}. This design can accelerate the scaled dot-product attention of Transformer and eliminate data dependency by avoiding writing the intermediate results using the matrix decomposition technique. For instance, Key $\mathcal{K}$ is the intermediate result from the computation of $\mathcal{K} = X \cdot W_K$, where $X$ is the input and $W_K$ is the weight matrix related to Key. Instead of computing with Key $\mathcal{K}$, the Query $\mathcal{Q}$ computes with the $W_K$ and $X$ successively. Each element in weight matrix is stored as an 8-bit number in four ReRAM cells, where each cell represents two bits. And the quantization-aware training can help maintain the inference accuracy of Transformer models and encoder-based Bert models~\cite{prato2019fully,zafrir2019q8bert}. 
Moreover, a hybrid computing unit for softmax computation is proposed, and a look-up table (LUT) is used for efficient exponential computation.
ReTransformer with 8-bit data representation improves computing efficiency by $23.21\times$ compared with GPU baseline. 

It is also worth noting that the normalization and activation calculations in Transformer involve nonlinear computations. To accelerate these computations, recent work NN-LUT proposes LUT-based neural network approximation to mimic the nonlinear computation with piece-wise linear computation~\cite{yu2021nn}. In addition, the CORDIC-based iterative method can also approximate the nonlinear computation and be implemented efficiently with IMC~\cite{truong2021racer}. The approximate computations can boost the computation efficiency without the sacrification of computing accuracy.


\section{Printed ML Circuits, An Ultra Resource Constrained Use Case}\label{sec:printed}

In this section, we use printed electronics as a use case example and investigate approximate printed ML circuits.
More specifically, we analyze how approximate computing can eventually enable, for the first time, the realization of battery-powered, high-accuracy, complex printed ML classifiers.
It is noteworthy that printed electronics form an ultra resource constrained environment and constitute an extreme scenario of embedded ML application.

\subsection{Printed Electronics Background}
While the Moore’s law has been the guiding force for the progress of lithography-based silicon Very Large Scale Integration (VLSI) technologies for higher integration density, these technologies have a lower bound on costs due to high costs of manufacturing (e.g., wafer processing, lithography, and material processing), which, in turn, increase the costs of packaging, testing, and assembly. 

\begin{figure}[t!]
\centering
\includegraphics[scale=0.30]{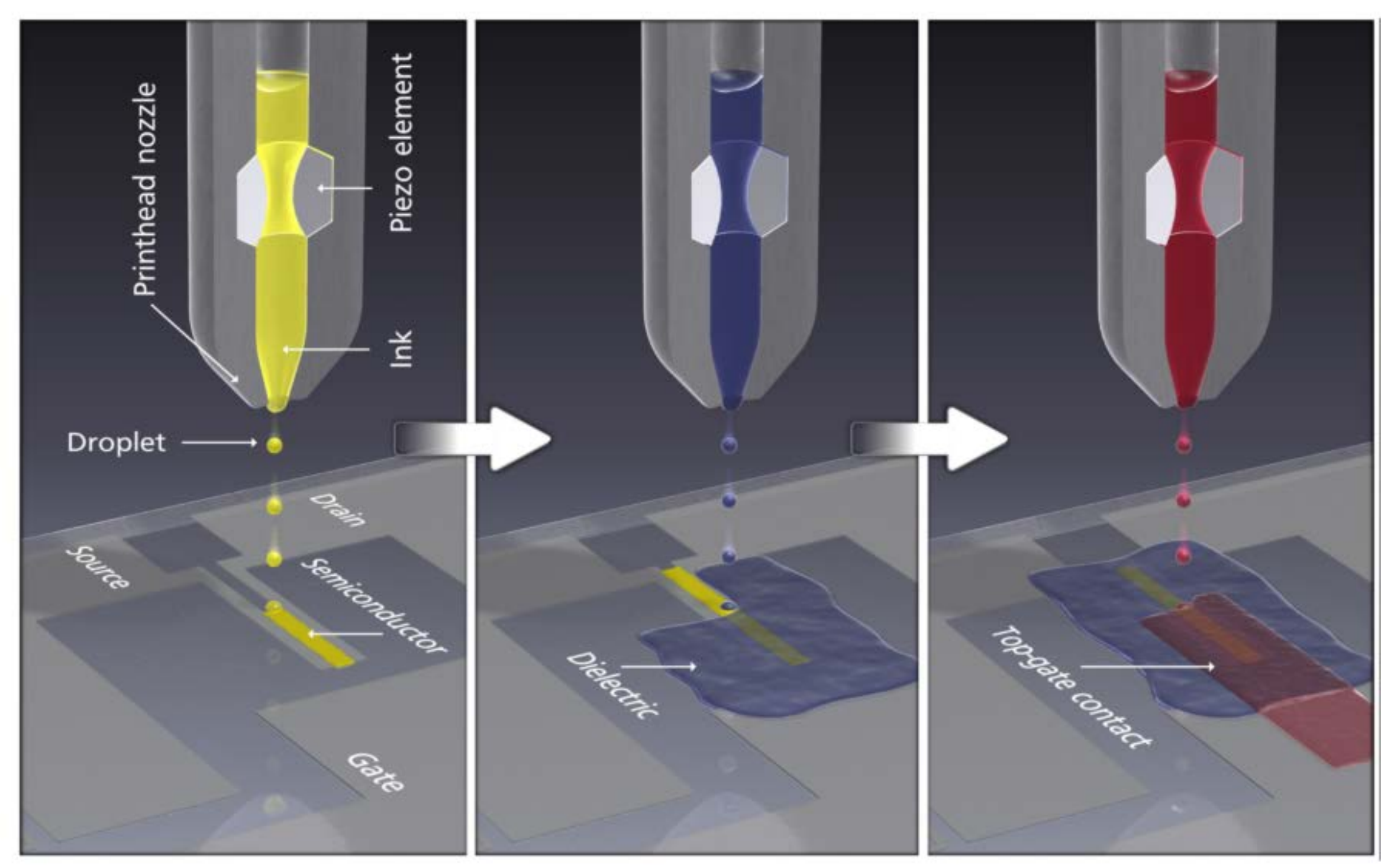}
\caption{Inkjet printing process}
\label{fig:pe}
\vspace{-2ex}
\end{figure}

An alternative manufacturing scheme, the so-called printed electronics, based on low-cost additive manufacturing technologies is a promising way to target disposables and ultra-low cost margin domains, especially with conformality needs. Printing technologies often rely on mask-less, portable, and additive manufacturing methods which can greatly reduce costs and production timelines~\cite{chang2017circuits}. 
Printed electronics refers to the fabrication technology that relies on printing processes, such as jet printing, screen or gravure printing~\cite{cui2016printed}.
Figure~\ref{fig:pe} shows the inkjet printing process of an electrolyte gated field-effect transistor (EGFET). 

Printed electronics technologies are broadly divided into two categories: 1) {\em additive manufacturing} process in which only deposition steps are involved, and 2) {\em subtractive process} involving a series of additive (deposition) and subtractive (etching) processes, similar to the silicon-based processing. 
The low equipment costs along with the simple additive manufacturing enable ultra low-cost (even sub-cent) electronic circuits.
Nevertheless, printed electronics cannot match the area and performance characteristics of silicon systems. The large feature sizes in printed electronics result in high device latencies and low integration density (orders of magnitude lower than in silicon VLSI). However, the applications in target domains typically have extremely low performance and precision requirements (e.g., sampling rate of few Hz and few bits precision) which may be met by printing technologies under acceptable area and energy constraints. 
Printed systems have been successfully fabricated, such as boolean logic~\cite{conti2020low} and a 32-bit Arm microprocessor~\cite{Biggs:Nature:2021}.

Due to its unique features, printed electronics constitutes a most-promising solution to eventually enable computing and bring intelligence in applications domains, that have not witnessed yet considerable
penetration of computing particularly in the 10-trillion market of  fast moving consumer goods (FMCG) \cite{lacy2020fast} such as smart packaging, low-end healthcare products, and disposables, to name a few. 
As a result, designing printed ML circuits has become the center of many research activities~\cite{Weller:ASPDAC:2020,Douthwaite:BIOCAS:2019,Ozer:Nature:2020,Mubarik:MICRO:2020:printedml} with remarkable results.
Nevertheless, despite the high efforts and the significant achievements reported, implementing complex printed ML circuits is still questionable through conventional computing.
The large hardware overheads in printed electronics mandate a higher degree of optimization that can be achieved only through alternative computing methods~\cite{Armeniakos:DATE2022:axml,Balaskas:ISQED2022:axDT,Weller:2021:printed_stoch}.

\begin{figure}[t!]
\centering
\includegraphics[scale=0.5]{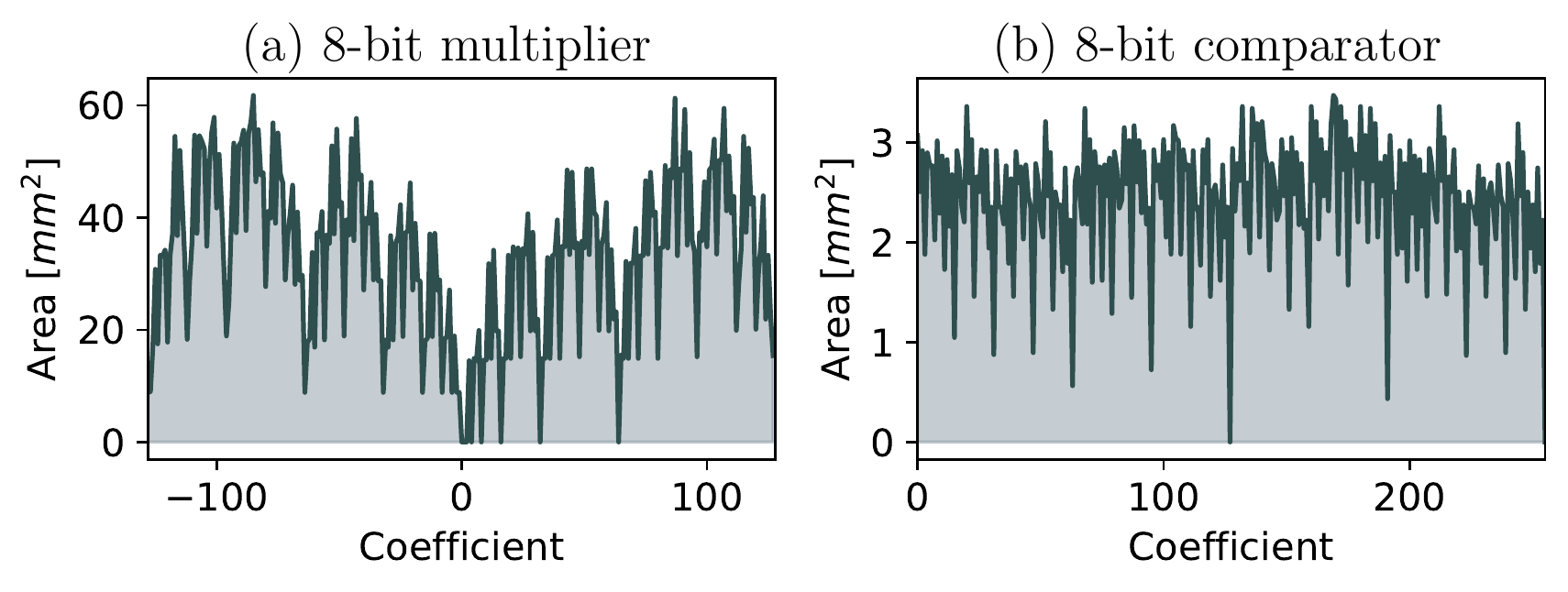}
\caption{Area variation of a bespoke (a) multiplier and (b) comparator with respect to the coefficient value~\cite{Armeniakos:DATE2022:axml,Balaskas:ISQED2022:axDT}}.
\label{fig:prcoefarea}
\vspace{-2ex}
\end{figure}

\subsection{Approximate Bespoke ML Classifiers}

The potential for large customization, that originates by the non-recurring engineering (NRE) cost and the low-cost in-situ and on-demand fabrication--even at low to moderate
fabrication volumes--of printed electronics, enables bespoke implementations~\cite{Mubarik:MICRO:2020:printedml,Bleier:ISCA:2020:printedmicro}.
In such bespoke circuits, the coefficients of the ML models are hardwired in the circuit implementation itself, leading to orders of magnitude lower hardware overheads (e.g., area, power) compared to conventional baseline implementations~\cite{Bleier:ISCA:2020:printedmicro}.
Exploiting the efficiency of a bespoke implementation and coupling it with approximate computing principles further reduces the hardware requirements paving the way, for the first time, towards complex, battery powered printed ML circuits~\cite{Armeniakos:DATE2022:axml,Balaskas:ISQED2022:axDT}.
It is noteworthy that bespoke circuits enable new types of approximation, allowing for more customized optimization, dedicated to each model.
In bespoke ML circuits the hardware overheads are highly correlated with the values of the coefficients of the model.
In other words, models with the same architecture but different coefficients will feature different hardware requirements.
For example, Figure~\ref{fig:prcoefarea} depicts the area overheads of 8-bit bespoke multipliers (Figure~\ref{fig:prcoefarea}(a)) and 8-bit comparators (Figure~\ref{fig:prcoefarea}(b)) for all coefficient values.
Note that multipliers and comparators constitute basic building blocks of most ML algorithms.
It is observed, in Figure~\ref{fig:prcoefarea}, that different coefficient values result in significantly different area. 
As a result, leveraging this property, by systematically approximating the coefficient values with hardware friendly ones, high area and consequently power gains (even up to 100\%) can be achieved.
In~\cite{Armeniakos:DATE2022:axml,Balaskas:ISQED2022:axDT} hardware-aware coefficient approximation is combined with traditional approximation techniques such as netlist pruning~\cite{ZervakisTVLSI2019:vader} and precision scaling.

Support Vector Machines (SVMs) and MultiLayer Perceptrons (MLPs) are targeted in~\cite{Armeniakos:DATE2022:axml}.
The core function of both SVMs and MLPs is to calculate a weighted sum.
Hence, by replacing each coefficient (in each neuron or classifier) with a more hardware-friendly value (e.g., from Figure~\ref{fig:prcoefarea}(a)) so that the positive errors (replace with smaller value) and the negative errors (replace with larger value) cancel out each other, high area savings can be achieved for a minimal accuracy loss.
This approximation is performed at the software level and the newly obtained model is used to generate the bespoke circuit.
Precision scaling is also leveraged to reduce the size of the required arithmetic operators.
Using only 4 bits for the inputs and 8 bits for the weights proved to barely impact the accuracy of all the models examined, delivering almost identical results to floating-point inference. 
Then, exploiting the fact that a significant portion of a circuit's gates switch very rarely, applying gate-level pruning at the hardware level (in which such gates are replaced by a constant 0 or 1) further increases the area gains.
Table~\ref{tab:pr_mlpsvm_gains} presents the hardware overheads of the exact and approximate classifiers when targeting a marginal accuracy loss of 1\%.
The datasets for training the models are obtained from the UCI ML repository~\cite{Dua:2019:uci}.
As shown, for all the examined models, employing approximate computing delivers large area (above 39\%) and power (above 34\%) gains.
What is even more important is that most of the approximated SVMs and MLPs can be powered by a printed battery (e.g., a Molex 30mW).
However, this is not the case for the exact bespoke baseline circuits.
Among the latter, only the RedWine MLP-R features adequate power consumption to be battery-powered.

\begin{table}[t]
\caption{Area and Power evaluation of approximate MLPs/SVMs for up to 1\% accuracy loss~\cite{Armeniakos:DATE2022:axml}.
}
\label{tab:pr_mlpsvm_gains}
\footnotesize
\centering
\setlength\tabcolsep{4pt}
\renewcommand{\arraystretch}{1.1}
\begin{threeparttable}
\begin{tabular}{l|ccc|cccc}
\hline
\multirow{2}{*}{\textbf{ML Circuit}} & \multicolumn{3}{c|}{\textbf{\begin{tabular}[c]{@{}c@{}}Exact\\ Bespoke\end{tabular}}} & \multicolumn{4}{c}{\textbf{\begin{tabular}[c]{@{}c@{}}Coef. Approx.\\ \& Gate Prune\end{tabular}}} \\ \cline{2-8}
\rule{0pt}{8pt}
& \begin{tabular}[c]{@{}c@{}}\textbf{Ac\tnote{\tiny 1}}\\  \end{tabular} &
\begin{tabular}[c]{@{}c@{}}\textbf{A\tnote{\tiny 2}}\\  \end{tabular} & \begin{tabular}[c]{@{}c@{}}\textbf{P\tnote{\tiny 3}}\\  \end{tabular} &
\begin{tabular}[c]{@{}c@{}}\textbf{A\tnote{\tiny 2}}\\  \end{tabular} & \begin{tabular}[c]{@{}c@{}}\textbf{P\tnote{\tiny 3}}\\  \end{tabular} & \textbf{\begin{tabular}[c]{@{}c@{}}AG\tnote{\tiny 4}\\ \end{tabular}} & \textbf{\begin{tabular}[c]{@{}c@{}}PG\tnote{\tiny 5}\\ \end{tabular}}  \\ \hline

\textbf{Cardio MLP-C} & 0.88 & 33 & 97 & 17 & 54 & 48\% & 44\%  \\
\textbf{RedWine MLP-C} & 0.56 & 18 & 53 & 8.0 & 27 & 55\% & 50\% \\
\textbf{WhiteWine MLP-R}  & 0.53 & 13.1 & 40.7 & 8.0 & 27 & 39\% & 34\% \\
\textbf{RedWine MLP-R} & 0.56 & 7.1 & 24 & 3.3 & 12 & 53\% & 49\%  \\
\textbf{RedWine SVM-C} & 0.57 & 24 & 73 & 7.6 & 26 & 68\% & 65\%\\
\textbf{Cardio SVM-C} & 0.90 & 15 & 47 & 8.7 & 29 & 43\% & 38\%  \\
\hline
\end{tabular}
\begin{tablenotes}\footnotesize
\item[] $^1$Accuracy. $^2$Area (cm$^2$). $^3$Power (mW). $^4$Area and $^5$Power Gain compared to the bespoke baseline.
\end{tablenotes}
\end{threeparttable}
\vspace{-6ex}
\end{table}

Similarly, printed Decision Trees (DTs) are examined in~\cite{Balaskas:ISQED2022:axDT}. 
Again, the authors exploit the area variation of the comparators with respect to the coefficient (threshold) value (Figure~\ref{fig:prcoefarea}(b)) and implement a software-based coefficient approximation. 
In addition, the circuit is further approximated/optimized by applying precision scaling at the input port of each bespoke comparator.
However, unlike~\cite{Armeniakos:DATE2022:axml}, the error due to coefficient approximation is not linear and its impact is hard to predict.
Therefore, in order to explore the space and identify for each comparator in the DT, the respective approximation configuration (i.e., input precision and value for the coefficient approximation), a non-dominated sorting genetic algorithm (NSGA-II) is used~\cite{Balaskas:ISQED2022:axDT}.
To guide the genetic algorithm to select more area efficient solutions, the accumulated area of the required comparators is used as a proxy of the DT’s area.
Table~\ref{tab:pr_dt_gains} presents the accuracy, area, and power evaluation of the exact and approximate DTs.
Again, the datasets are obtained from the UCI ML repository~\cite{Dua:2019:uci}.
As shown in Table~\ref{tab:pr_dt_gains}, the approximate DTs achieve more than 47\% area and more than 48\% power gains.
Most importantly, all the approximate printed DTs can be powered by a small Blue Spark printed battery since they feature less than 3mW power consumption.
In addition, the approximate Seeds can also be self-powered by an energy harvester as its power consumption is less than 0.1mW.
Finally note that the area of the majority of the approximate DTs is well constrained and as a result they are perfect candidates for printed applications.


\section{Lessons Learned and Open Issues}
Approximate computing has been considered in a variety of application domains as an effective alternative in building efficient computing systems.
Though, approximate computing exploitation really took off in ML.
As widely demonstrated in the state of the art~\cite{axdnnsurv} and presented by several use cases in this paper, approximate computing and ML form a perfect synergy and their matching not only feels natural, but it seems mandatory in order to address the increased computing challenged in ML systems. 
ML models which try to approximate and learn the functions, naturally fit with the concept of approximate computing which, under bounded accuracy impacts, try to further reduce the hardware footprint (area, delay, power)~\cite{AxNN,rapid} or even address thermal constraints~\cite{axthermal,nputhermal} and/or defend against advertasial attacks~\cite{Panda:ACCESS2019,Guesmi:ASPLOS2021}

\begin{table}[t]
\caption{Accuracy, Area, and Power evaluation of approximate Decision Trees~\cite{Balaskas:ISQED2022:axDT}.
}
\label{tab:pr_dt_gains}
\footnotesize
\centering
\setlength\tabcolsep{3pt}
\renewcommand{\arraystretch}{1.1}
\begin{threeparttable}
\begin{tabular}{l|ccc|ccccc}
\hline
\multirow{2}{*}{\textbf{ML Circuit}} & \multicolumn{3}{c|}{\textbf{\begin{tabular}[c]{@{}c@{}}Exact\\ Bespoke\end{tabular}}} & \multicolumn{4}{c}{\textbf{\begin{tabular}[c]{@{}c@{}}Coef. Approx.\\ \& Precision Scaling\end{tabular}}} \\ \cline{2-9}
\rule{0pt}{8pt}
& \begin{tabular}[c]{@{}c@{}}\textbf{Ac\tnote{\tiny 1}}\\  \end{tabular} &
\begin{tabular}[c]{@{}c@{}}\textbf{A\tnote{\tiny 2}}\\  \end{tabular} & \begin{tabular}[c]{@{}c@{}}\textbf{P\tnote{\tiny 3}}\\  \end{tabular} &
\begin{tabular}[c]{@{}c@{}}\textbf{Ac\tnote{\tiny 1}}\\  \end{tabular} &
\begin{tabular}[c]{@{}c@{}}\textbf{A\tnote{\tiny 2}}\\  \end{tabular} & \begin{tabular}[c]{@{}c@{}}\textbf{P\tnote{\tiny 3}}\\  \end{tabular} & \textbf{\begin{tabular}[c]{@{}c@{}}AG\tnote{\tiny 4}\\ \end{tabular}} & \textbf{\begin{tabular}[c]{@{}c@{}}PG\tnote{\tiny 5}\\ \end{tabular}}  \\ \hline

\textbf{Arrhythmia} & 0.56 & 163 & 7.6 & 0.67 & 22 & 1.04 & 86\% & 86\% \\ \hline
\textbf{Balance} & 0.75 & 68 & 3.1 & 0.81 & 27 & 1.16 & 60\% & 63\% \\ \hline
\textbf{HAR} & 0.84 & 551 & 26.1 & 0.83 & 295 & 13.7 & 47\% & 48\% \\ \hline
\textbf{Mammographic} & 0.76 & 99 & 4.5 & 0.81 & 8.06 & 0.38 & 92\%  & 92\% \\ \hline
\textbf{Seeds} & 0.89 & 30 & 1.4 & 0.94 & 2.32 & 0.09 & 92\% & 94\% \\ \hline
\textbf{Vertebral} & 0.85 & 58 & 2.7 & 0.86 & 7.84 & 0.38 & 86\% & 86\% \\ \hline

\hline
\end{tabular}
\begin{tablenotes}\footnotesize
\item[] $^1$Accuracy. $^2$Area (cm$^2$). $^3$Power (mW). $^4$Area and $^5$Power Gain compared to the bespoke baseline.
\end{tablenotes}
\end{threeparttable}

\end{table}

Among a considerable amount of varying approximations for ML that have been investigated, precision scaling establishes as the most prominent one, leading to an actual success story.
ML models feature such a high degree of resilience that can tolerate very aggressive precision scaling, down to only a few bits~\cite{Choi:2018}.
It is noteworthy that many commercial ML and DNN accelerators, such as GPUs, FPGAs, and TPUs, have already adopted precision scaling (e.g., from floating point to low precision fixed point arithmetic) as well as quantized neural networks (QNNs) have become a norm~\cite{swagath2020,rapid}.
This further highlights the fact that some of the core concepts of approximate computing have been well embraced by the ML community.
One of the reasons that precision scaling shined in ML is that it features a straightforward design and exploitation and preserves the regularity of compute.
In other words, it delivers very high (and highly correlated) gains in both the compute, the memory, and the data transfer, not affecting, thus, the application’s bottleneck. 
Moreover, it maintains the level of design abstraction and the link between software and hardware is apparent, leading also to straightforward quantification of the induced error.
On the other hand, this is not the case for other approximation techniques (e.g., logic) in which the impact of approximation is irregular and require extremely slow hardware emulation~\cite{axadapt} or circuit simulation~\cite{axweight}.
Finally, as discussed in Section~\ref{sec:overview}, to exploit the full potential of approximation, proper hardware support is required (e.g., architecture support for variable precision and/or sparsity).
On the other hand, this is not the case in bespoke implementations, as for example in printed ML circuits and fully parallel FPGA architectures.
Bespoke architectures (Section~\ref{sec:printed}) exploit out-of-the-box all the benefits that may originate from low precision inputs/weights and pruning.
In bespoke ML classifiers, unstructured pruning will directly result in eliminating the respective multipliers and reducing the size and potentially the precision of the accumulation tree.
Moreover, low-precision inputs or small-value weights will have the same impact, i.e., significantly reduced hardware overheads for multiplications and additions.
Hence, such features can be additionally exploited for further approximate ML models in customized hardware realizations.

Still, despite the early emplacement of approximate computing in ML, there are still several key challenges that need to be addressed in order to enable further its growth and adoption.
One of the major open research challenges of approximate computing, particularly when combined with machine learning, is the verification, i.e., how to quantify and bound the accuracy loss due to approximation, particularly in the context of mission critical tasks and applications.
Related to verification challenges, it is important to quantify (and prove) the extent of uncertainty and inaccuracy that is imposed by an approximate computing technique for a specific model since approximation adds another degree of inaccuracy on top of ML models which are inherently inexact.
An open challenge is to understand what types of approximations can still be utilized, especially for critical applications, and what are the limits of approximate computing in such domains.
Moreover, the impact of approximate computing on the robustness of the ML model is merely explored. 
Approximate techniques aim mainly in improving performance and/or reducing energy by removing, however, redundancy from the ML model and/or by inducing noise in the computations (e.g., approximate units).
A reasonable assumption would be that the model becomes less robust and, for example, more vulnerable to memory errors.
Nevertheless, this trade-off hasn't been comprehensively analyzed yet.
Finally, a key challenge refers to the design and optimization of approximate ML accelerators.
While approximate computing can further simplify design closure and reduce hardware footprint, one challenge is to ensure that this does not lead to an explosion of the design space since now the choice of approximation adds additional decision dimensions.
In addition, re-training/approximation-aware training/quantization-aware training are widely and mainly used to mitigate the accuracy degradation due to approximation.
Still, re-training (especially in DNNs) can be time consuming, increasing the design cycle time and exacerbating the complexity of the associated design space exploration.
Overall, how to approximate the design choices for approximate computing seem to be a very promising research direction in design automation for approximate computing in the context of machine learning.

\begin{acks}

This work is partially supported by the German Research Foundation (DFG) through the project ``ACCROSS: Approximate Computing aCROss the System Stack'' HE 2343/16-1 and by grant from the Excellence Initiative of Karlsruhe Institute of Technology under Future Field program ``SoftNeuro''. This work is partially supported by US NSF 1955196, 2112562, 1910299, 1937435, and ARO W911NF-19-2-0107.

\end{acks}

\printbibliography

\end{document}